\input cp-aa.tex

%
  \MAINTITLE{Boron in the extreme Pop~II star HD\,140283 and the
production of light elements in the Early Galaxy\FOOTNOTE
{Based on observation with the NASA/ESA Hubble Space Telescope,
obtained at the Space Telescope Science Institute, which is operated by the
Association of Universities for Research in Astronomy, Inc., under NASA
contract
NAS5-26555.}
}
  \AUTHOR{
B. Edvardsson@1, B. Gustafsson@1, S.G. Johansson@2@,@3, D. Kiselman@4,
D.L. Lambert@5, P.E. Nissen@6, G. Gilmore@7
}         
  \INSTITUTE={
@1 Astronomical Observatory, Box 515, S-751 20 ~Uppsala, Sweden
@2 Department of Physics, University of Lund, S\"olvegatan 14, S-223 62 ~Lund,
Sweden
@3 Lund Observatory, University of Lund, Box 43, S-221 00 ~Lund, Sweden
@4 NORDITA, Blegdamsvej 17, DK-2100 K\o benhavn \O, Denmark
@5 Department of Astronomy, University of Texas, Austin, TX 78712-1083, U.S.A.
@6 Institute of Physics and Astronomy, University of Aarhus, DK-8000 Aarhus C,
Denmark
@7 Institute of Astronomy, Madingley Rd., Cambridge CB3 0HA, England
}      
  \ABSTRACT={
Using observations of the 2496.7\,\AA\ B\,{\sc i} line with the HST GHRS at a
nominal resolution of 90,000, we have found the abundance of boron of
HD\,140283
to be $\log \epsilon_{\rm B} (= 12 + \log (N_{\rm B}/N_{\rm H}))=0.34 \pm
0.20$.
This result is found when a significant non-LTE effect in the formation of the
B\,{\sc i} line is taken into account.
The resulting $N_{\rm B}/N_{\rm Be}$ ratio is about 17 (in the range 9 -- 34),
which is in very good agreement with what is expected from spallation by cosmic
rays.
We conclude that this origin of Be and B in the Early Galaxy is the most
probable of recently suggested formation mechanisms.
}
  \KEYWORDS={ Galaxy: abundances -- Stars: abundances -- Line: formation }
  \THESAURUS={7 (10.01.1; 08.01.1; 02.12.1)
 }      
  \OFFPRINTS={ Bengt Edvardsson }      
  \DATE={ ????? }           
%
\maketitle
  \MAINTITLERUNNINGHEAD{Boron in HD\,140283}
  \AUTHORRUNNINGHEAD{Edvardsson et~al.}

\titlea{Introduction}
The detection of beryllium in the Extreme Population II star HD\,140283
([Fe/H]$\;\approx-2.7$) by Gilmore, Edvardsson \& Nissen (1991) indicated a
value of $\log({\rm Be/H})$ of about $-13$, a factor of more than 1000
times greater than the primordial value predicted by the Standard Big Bang
Model.  (Here and below the notation Be/H is shorthand for
${N_{\rm Be}/N_{\rm H}}$.)
This relatively high abundance was not expected, and it is still not quite
clear
where and how the beryllium was formed: in the interstellar medium (ISM) in the
Early Galaxy, in supernovae or their immediate
neighbourhood, close to other high-energy objects, or possibly in the Big Bang.

Although it seemed that the Be abundances could indicate an inhomogeneous
Big Bang (see, e.g., Kajino \& Boyd 1990) this possibility was questioned on
theoretical grounds (e.g., Terasawa \& Sato 1990, Thomas et~al. 1993 and
references given therein).
Also, the approximately linear relation between Be and O abundances found for
Pop~II dwarfs by Gilmore et~al. (1992) suggested a non-cosmological origin of
the Be in HD\,140283 and a gradual build-up of the element in the Halo.
However, the relation found deviates considerably from the quadratic behaviour
of single-zone models for spallation in the ISM of the Early Galaxy
(cf. Vangioni-Flam et~al. 1990).
This has lead to several modified scenarios for cosmic ray spallation in the
Early Galaxy (for references, see Sect.~4).
Neutrino-induced spallation in supernovae may contribute significantly to some
light-element abundances (Woosley et~al. 1990) and possibly also to Be
(Malaney 1992).
A fourth possibility is that a fraction of the beryllium in HD\,140283 was
produced by other (pregalactic) processes (e.g. by ``photo erosion'' at an
early active galactic nucleus, Boyd \& Fencl 1991).

Duncan, Lambert \& Lemke (1992) searched for the B\,{\sc i} 2496.7\,\AA\ line
in the spectrum of HD\,140283, using Hubble Space Telescope (HST) observations
at a spectral resolution of about 25,000.
The line was tentatively identified and an abundance of about
$\log({\rm B/H}) = -12.1$ was estimated by an LTE abundance analysis.
The resulting abundance ratio of B/Be$\;= 10$ is consistent with spallation;
however, this result is very uncertain and the ratio may sooner be regarded a
rough upper limit since the line observed is a blend of three lines,
Co\,{\sc i}, B\,{\sc i} and Fe\,{\sc i}, which could not be resolved at the
spectral resolution used.
Thus, observations at considerably higher spectral resolution and high signal
to noise were needed.
Here, we present such observations in Sect.~2.
The spectra are analysed in Sect.~3 and the results are discussed in Sect.~4.

\titlea{Observations and data reductions}
\titleb{Observations}
The wavelength region around the resonance lines of B\,{\sc i} at
2497\,\AA\ was observed with the
Goddard High Resolution Spectrograph (GHRS) of the Hubble Space Telescope
on September 5, 1992 (proposal No. 3479) and continued on February 15 and 21,
1993 (proposal extension No. 4766).
A description of the spectrograph and general observational techniques may be
found in the manual by Duncan (1992).
The target was found each time without any reported problem by the standard
onboard target acquisition procedures.
The Echelle~B grating and the Small Science Aperture (SSA), 0\farcs 25 by
0\farcs 25, was used which optimally enables a resolving power
$R={{\lambda}/{\Delta\lambda}}=90,000$.
A spectral range of about 12\,\AA\ centered near $\lambda_{\rm vac}=2494$\,\AA
A sub-stepping pattern (No.~7) of 4 times $1/4$ diode steps was used to
fully sample the spectrum projected on the 500 science diodes to obtain the
highest resolution.
In order to enable background subtraction we used 6\% of the exposure time to
expose the Echelle inter-order regions on both sides of the spectrum.
A typical set of 4 sub-steps and 2 background observations of Sept. 1992 are
shown in Fig.~1.
\begfig 1.0cm
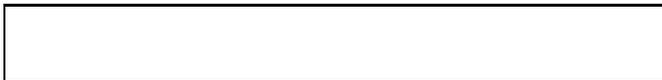
\figure{1a-f}{
4 sub-step exposures on HD\,140283, and 2 background observations.
Mean count levels are 1.5 counts per diode for the stellar spectra and 0.03
counts for the background.
The total exposure time was 81.6 seconds with 94\% of the time spent for the
stellar spectra
}
\endfig
The observations were performed as many short integrations to avoid as much as
possible the GIMP (geomagnetical image motion problem), which may move the
spectrum across the Digicon detector and degrade the spectral resolution.
Furthermore, to minimise the impact of diode-to-diode sensitivity differences,
the spectra were obtained at four different detector positions (FP-SPLIT=4).

The 1992 data were acquired during 9 spacecraft orbits with all exposures
for each FP-split position taken consecutively (first 65 exposures in the first
position, then 65 in the second position etc.).
The S/N ratio of this observation turned out to be lower than what was hoped
for, why the rest of the observing time allotment was utilized for the same
star with small modifications (which gave the programme a new identification
number).
For the 1993 observations, the procedure was altered somewhat in an attempt to
improve the relative wavelength calibration between separate orbits.
One wavelength calibration spectrum (Pt-Ne lamp) was obtained in the
beginning of each spacecraft orbit.
This procedure resulted in 7 spacecraft orbits being used for each of the
two observations in 1993.
The 4 FP-split observations were this time taken consecutively for each
exposure.

The summed exposure times (on star) for each date were
5$^h$32$^m$23$^s$; Sept. 5, 1992; 65 exposures
(each consisting of 4 FP-splits $\times$ 4 sub-steps),
2$^h$58$^m$59$^s$; Feb. 15, 1993; 35 exposures and
2$^h$40$^m$38$^s$; Feb. 21, 1993; 32 exposures.
The reported continuum fluxes at the three occasions were, respectively,
$0.69\cdot 10^{-12}$, $0.85\cdot 10^{-12}$ and
$1.88\cdot 10^{-12}$\,erg\,cm$^{-2}$\,s$^{-1}$\,\AA$^{-1}$.
We attribute these differences to pointing errors with the small aperture.
If the highest flux level had been obtained at all three occasions, 85\% more
photons would have been registered, corresponding to 36\% higher a S/N ratio
of the final spectrum.
Thanks to improved onboard acquisition procedures ("SSA peak-up") we
believe that such pointing errors are now avoidable.

\titleb{Data reductions}
The data reductions were performed using the IRAF/STS\-DAS software under
guidance of Dr.\,Jeremy Walsh at the European Coordination Facility (ECF) in
Garching bei M\"unchen, Germany, in October 1992 and September 1993.
The most important aim of the data reductions was to wavelength align the very
large number of short exposures as well as possible before adding them to a
final spectrum.
Each short exposure has a very low S/N ratio (cf. Fig.~1), and slow,
unpredictable drifts in the wavelength direction make their co-addition
difficult.
Any mis-alignment will cause a degradation of the high spectral resolution
sought in this project.

As a starting point for the reductions we used the standard STScI processed
files (this processing procedure is called ``pipelining'').
The pipeline files contain the combinations of each set of four 500 diode
sub-step spectra to a 2000 channel fully-sampled spectrum, with the background
subtracted and converted from counts to flux units.
The pipelining takes into account: correction fore individual diode responses,
vignetting, background subtraction, echelle blaze function removal and the
conversion to absolute flux.
The conversion to absolute flux can, however not take telescope pointing into
account, why the resulting absolute fluxes are lower limits to the true
absolute fluxes.
Preliminary wavelength scale files are also provided by the pipeline process.

We used the STSDAS version 1.1 for the reductions of the 1992 data.
We aligned the 4 FP-SPLIT files for each of the 65 exposures, which resulted in
65 separate spectra.
Their relative wavelength offsets were determined by auto-correlation
and then added together.
As an alternative, we attempted to use the recorded FP-SPLIT wavelengths
with a correction for the GIMP effect as determined from the relevant
component of the geomagnetic field in the GHRS for each integration.
The method to extract the relevant data was devised by Dr.\,J.\,Walsh, but in
this case it did not result in any discernible improvement over the standard
STSDAS procedures.

For the 1993 data we used version 1.2 of STSDAS.
We attempted to use the wavelength calibration files obtained during
each orbit to guide in the spectral alignment, but this did not significantly
improve the result.
The spectra resulting from the three observing dates are shown in Fig.~2.
\begfig 1.0cm
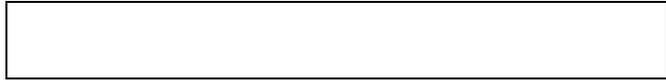
\figure{2a-c}{
The summed spectra obtained at our 3 observing dates before wavelength
normalisation and continuum rectification.
The dotted lines show the estimated continuum position
}
\endfig

The channel width is 5.9506\,m\AA\ and from the variance of the
channel-to-channel flux we estimate the continuum S/N
ratio per channel of the final spectrum, shown in Fig.~3, to be about 33.
This is close to the statistical S/N ratio expected from the about 1400
counts per channel accumulated in the continuum.
\begfig 1.0cm
\figure{3}{
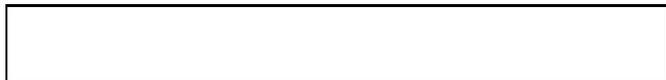
The final spectrum after wavelength and continuum normalisation
}
\endfig
{}From our theoretical models, the continuum level of HD\,140283 was estimated
to increase by 1.8\% per 10\,\AA\ at 2500\,\AA.
A good continuum window appears around $\lambda_{\rm obs}=2498.8$ (in Fig.~2,
compare also with Fig.~1 of Duncan et~al. 1992).
Fixing a straight line through the noise at this point to the standard flux
calibrated spectra with a slope of 1.8\% per 10\,\AA\ gives a very reasonable
continuum placement, cf. Figs.~2 and 3.
We adopt this continuum level and estimate a maximum uncertainty of 2\% in the
continuum.
The resulting line blocking in this wavelength region is 16.5\% for HD\,140283.

In Fig.~4 we compare the equivalent widths of several absorption features
in our spectrum with those measured from the lower resolution GHRS spectrum of
HD\,140283 by Duncan et~al. (1992).
For 11 lines between 2491 and 2495\,\AA\ there is a systematic difference of
6.7\,m\AA\ with a scatter of 5.9\,m\AA (Duncan et~al. minus us).
This is probably caused by a lower continuum placement in this region in our
spectrum.
For 13 lines in the wavelength region containing the B\,{\sc i} lines,
2495 -- 2499\,\AA, the equivalent widths agree very well: the difference is
$1.2\pm 3.1$\,m\AA.
\begfig 1.0cm
\figure{4}{
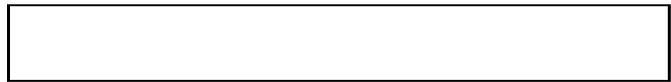
Comparison of our equivalent widths with those of Duncan et~al. (1992).
The line shows loci for identical equivalent widths.
}
\endfig

\titlea{The boron abundance}
\titleb{LTE abundance analysis and error estimates}
The abundance analysis was based on the methods used in Gilmore et~al. (1992).
The theoretical model atmospheres were described in Edvardsson et~al. (1993).
The effective temperature, 5680\,K, and surface gravity $\log g=3.5$ was
adopted
from Nissen (1993) who estimates an interstellar reddening $E_{b-y}=0.020$ for
HD\,140283 based on $uvby\beta$ photometry and interstellar Na\,{\sc i} D-line
absorption.
Our recent analyses (Gilmore et~al. 1991, 1992; Nissen et~al. 1993) based the
effective temperature estimate of this star on the assumption that the star is
unaffected by interstellar reddening.
The overall metallicity [Fe/H]$\;=-2.64$ was adopted from Nissen et~al. (1993),
with a modification due to our 140\,K higher $T_{\rm eff}$.
The microturbulence parameter was taken to be $\xi_t=1.5$\,km\,s$^{-1}$.
These atmospheric parameters are very similar to values used by
Duncan et~al. (1992) and other recent work on this star.

The line list was taken from Duncan et~al. (1992), with a modified wavelength
of
2496.716\,\AA\ (Pickering 1993) for the Co\,{\sc i} line which blends with the
primary B\,{\sc i} line at 2496.772\,\AA.
We consider the line identifications to be reliable and, in particular,
we find it very unlikely that there should be other blending lines than the
Co\,{\sc i} line that contribute significantly to the 2496.772\,\AA\ feature,
i.e. it is very unlikely that the observed feature is not the boron line.
The oscillator strengths for all except the B\,{\sc i} lines were adjusted to
fit the lines observed in the GHRS spectrum.
Adjustments of the $gf$ values were, however, usually unnecessary.
In particular, the Co\,{\sc i} 2496.716\,\AA\ line and the Fe\,{\sc i}
2496.870\,\AA\ line closest to the 2496.772\,\AA\ line were not changed.
To account for the combined effects of macroturbulence, rotational broadening
and instrumental profile the synthetic spectrum was convolved with a Gaussian
profile with a FWHM of 5.5\,km\,s$^{-1}$ (45\,m\AA).
Our best fit synthetic spectrum is shown in Fig.~5, and corresponds to an LTE
boron abundance of
$\log \epsilon_{\rm B} (=12 + \log{{N_{\rm B}}\over{N_{\rm H}}})$ of $-0.20$.
In panel {\bf b} the observed spectrum was convolved by a three-point
triangular
profile of the shape 0.25, 0.50, 0.25.
\begfig 1.0cm
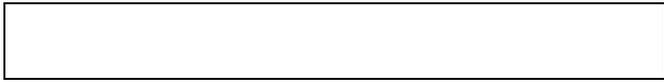
\figure{5a-b}{
Comparison of observed (solid lines) and synthetic spectra calculated in LTE
(dashed lines).
In {\bf b} the observed spectrum has been somewhat smoothed.
The short-dashed line shows our best synthetic spectrum,
$\log \epsilon_{\rm B}{\rm (LTE)}=-0.20$ and the long-dashed lines are
synthetic
spectra with $\pm 0.3$\,dex varied boron abundances.
The dotted line shows the synthetic spectrum with only the boron line,
which has an equivalent width of 5.0\,m\AA.
The boron line is about $2/3$ as strong as was suggested in the lower
resolution
study of Duncan et~al. 1992 (cf. their Fig.~9)
}
\endfig

The other line of the B\,{\sc i} doublet, $\lambda=$2497.723\,\AA, has twice as
large transition probability as the 2496.772\,\AA\ line, but since it is
blended by stronger Fe\,{\sc i} and Fe\,{\sc ii} lines with poorly known
transition probabilities it is currently not useful as a reliable boron
abundance indicator.

To explore the sensitivity of our LTE boron abundance to various sources of
uncertainties, detailed synthetic spectra were calculated for several
combinations of model atmosphere fundamental parameters and abundances.
When the effective temperature of the model was lowered by 100\,K the
resulting boron abundance decreased by 0.11\,dex.
The sensitivity to the surface gravity and microturbulence parameters were
small: $\Delta \log \epsilon_{\rm B} / \Delta \log g =-0.02 / +0.30$ and
$\Delta \log \epsilon_{\rm B} / \Delta \xi_t =0.00 / +0.50$\,km\,s$^{-1}$.
The 2\% maximum uncertainty of the continuum level introduces an rms abundance
uncertainty of $\pm0.10$\,dex.
The statistical uncertainty in the boron line flux due to photon noise
(continuum S/N$\;\ga 33$ per channel and about 10 channels dominated by the
B\,{\sc i} line at typically 95\% of the continuum level) is about
$\la \pm 1$\%, which corresponds to $\la \pm 20$\% in the equivalent width or
$\la 0.10$\,dex in abundance.
The adjustment of the wavelength of the blending Co\,{\sc i}
$\lambda 2496.716$\,\AA\ line was +0.009\,\AA, which, due to the high spectral
resolution, does not appreciably affect the resulting boron abundance.
We have checked the profile of the Co\,{\sc i} line observed in the laboratory
(Pickering 1994) for hyperfine structure (hfs) and find that there is an
unresolved pattern around the center of gravity wavelength of 2496.716\,\AA.
The major hfs components lie, however, only $\pm 6$\,m\AA\ from the
center of gravity wavelength and the effect of the Co\,{\sc i} hfs on
the boron line is therefore negligible.
Nor does the uncertainty in the $gf$ value of the Co\,{\sc i}
line introduce any noticeable uncertainty in the boron abundance.

\titleb{Non-LTE line formation}
Kiselman (1994) investigated the statistical equilibrium
of neutral boron in solar-type stars and found that significant
departures from LTE are expected in a metal-poor star such
as HD\,140283. The departures consist of an overionisation effect
driven by the optical pumping in the ultraviolet resonance lines
together with the pumping in the observed 2496.7\,\AA\ line itself.
The overionisation decreases the line opacity and the optical
pumping raises the line source function above the Planckian value.
These effects combine to make the line weaker than in LTE.
An LTE analysis will thus underestimate the boron abundance.

We have made non-LTE calculations for the current model atmosphere using the
same methods and the same atomic model as in Kiselman (1994).
The result is that the LTE abundance should be increased with $+0.54$\,dex.
The main uncertainty in this figure comes from the effect of blending by other
lines in the B\,{\sc i} resonance lines which causes a moderation
of the optical pumping effect of unknown magnitude.
Most of the atomic data is of considerable accuracy and
should cause smaller errors.
{}From numerical experiments with the adopted model atmosphere, we estimate the
maximal allowed range of the non-LTE abundance correction to be between
$+0.30$\,dex and $+0.60$\,dex.
The non-LTE effects also cause greater sensitivity
of the equivalent width to the atmospheric parameters.
The effect increases in magnitude with $0.07$\,dex when $T_{\rm eff}$ is
increased by 100\,K and decreases with $0.09$\,dex when $\log g$ is
increased by 0.5\,dex (Kiselman 1994).

Beryllium has not yet been the subject of an extensive non-LTE
investigation. If the same general mechanisms as in the
boron case should apply also for beryllium, we may expect
that the overionisation of Be\,{\sc i} and the optical pumping in the observed
ultraviolet Be\,{\sc ii} line will tend to cancel the effect of each other
in the abundance estimate from Be\,{\sc ii}.
We therefore expect that the departures from LTE will be small for the
observed 3131\,\AA\ line. Indeed, preliminary
calculations using a 93-level model shows that the equivalent
width of the 3131\,\AA\ is only 0.06\,dex weaker than in LTE.
It seems reasonable to assume that the non-LTE effect for
the beryllium abundance is $0.0 \pm 0.2$\,dex (maximum error).

\titlea{Discussion}
The B/Be abundance ratio depends on the origins of the two elements, as
discussed below, and is therefore a valuable clue to their formation.
The beryllium abundance of HD\,140283 was determined by
Gilmore et~al. (1991 and 1992) from the Be\,{\sc ii} resonance
doublet at 3131\,\AA.
The weighted mean result given by the latter investigation was
$\log \epsilon_{\rm B}=-0.97 \pm 0.25$, which,
when scaled to our effective temperature, is translated to $-0.90 \pm 0.25$.
It should be mentioned here that Ryan et~al. (1992) used spectra of somewhat
lower spectral resolution and their result for HD\,140283 corresponds to
$\log \epsilon_{\rm Be}=-1.13 \pm 0.4$ at our surface gravity.

The abundance ratio $N_{\rm B}/N_{\rm Be}$ found in this study is 17, when
corrections (+0.5\,dex for B; 0.0\,dex for Be) for departures from LTE have
been
applied.
What errors can be ascribed to this ratio?
The observational errors in B abundance amount to about 0.14\,dex (rms); those
due to errors in the fundamental parameters for HD\,140283 to 0.12\,dex, and
those resulting from errors in the non-LTE analysis to 0.1\,dex.
For the Be abundance, the corresponding errors are 0.17\,dex, 0.11\,dex
and 0.1\,dex, respectively.
Taking the correlated effects of parameter uncertainties into consideration,
and adding the abundance errors quadratically (in linear measure) we find the
rms error in the ratio to be 0.3\,dex.

The resulting B/Be ratio is in the range 9--34 with 17 being the most probable
value.
This is in very good agreement with predictions for cosmic ray spallation.
Thus, the ratio 14.5 was predicted for spallation in the
halo gas (with [O/Fe]$\;=+0.5$ and $\alpha/p=0.08$) by Steigman \& Walker
(1992)
with a cosmic ray energy spectrum as that of the present day solar
neighbourhood.
In order to produce a ratio significantly above 20, say, a spectrum strongly
peaked around 40\,MeV would be needed, but even in such a rather contrived case
a B/Be ratio above 30 is hardly probable (cf. Duncan et~al. 1992, Fig.~5).
The cross sections for the production of Be and B are almost
independent of energy for higher energies than 100\,MeV/nucleon and thus the
production ratio stays constant and independent of the details of the
cosmic ray spectrum as long as it is peaked towards high energies.
Duncan et~al. (1992) calculated the high energy asymptotic ratios from 10 to 12
depending on the C/Fe ratio and suggested a lower limit of B/Be$\;= 7$ when the
uncertainties in cross sections are taken into consideration.
A cosmic ray spectrum heavily peaked towards energies below 8\,MeV/nucleon may
result in a significantly greater B/Be ratio, due to the threshold for Be
production by cosmic ray alpha particles at that energy.
However, obviously our measured B/Be ratio for HD\,140283 is in
very good agreement with what one would expect from production through
cosmic ray spallation as long as the CR spectrum is not very strongly
peaked around 40\,MeV/nucleon or towards very low energies.

Is it safe from this result to conclude that the beryllium and boron in
Population II stars were indeed formed by cosmic-ray spallation?
Alternative production mechanisms include neutrino-induced production in
supernovae (cf. Woosley et~al. 1990, Malaney 1992, Olive et~al. 1993),
"non-standard" Big-Bang models (Boyd \& Kajino 1989, Malaney \& Fowler 1989,
Kajino \& Boyd 1990)
or photoerosion near active galactic nuclei or similar objects
(Boyd \& Fencl 1991).

Malaney (1992) estimated that the lowest B/Be ratio allowed for by the
neutrino-induced SN production may be about 30, but the proposed Be production
mechanism is intricate and Malaney cautioned that more detailed
models of metal-poor supernovae are needed to verify this estimate.
More recent models of Type~II supernovae with neutrino-induced spallation by
Timmes, Woosley \& Weaver (quoted by Olive et~al. 1994) seem to lead to
significantly higher B/Be ratios.
Also, Olive et al. (1993) found a high value ($\ga 40$) for a galactic
evolution
model in which both neutrino processes and spallation nucleosynthesis were
included in an attempt to reproduce the solar $^{11}$B/$^{10}$B ratio.

The inhomogeneous Big-Bang models tend to give much lower ratios than 10;
e.g. Boyd \& Kajino (1989) obtained one order of magnitude less B than Be in
their models.
However, there may be exceptions to this (cf. the low Be abundances obtained
in some models by Malaney \& Fowler 1989, see also Kawano et~al. 1991).
More important is, however, that the production of light elements according to
more recent and sophisticated models
result in yields of B and Be orders of magnitude below those found earlier.
These models include the vital diffusion of neutrons and protons
before, during and after nucleosynthesis and contain more extensive reaction
networks than used earlier
(cf. Thomas et~al. 1993, Jedamzik et~al. 1994, and references quoted therein).
It seems from this that inhomogeneous Big Bang nucleosynthesis, at least in the
form currently conceived, is not responsible for the Be and B abundances found
in HD\,140283.

The photoerosion network calculations by Boyd \& Fencl (1991) give B/Be ratios
$\ga 10$, with low values only for gamma ray fluxes great enough, or
irradiation times long enough for quasi-equilibrium to be nearly established.
E.g., we estimate from Fig.~1 of Boyd \& Fencl (1991) that an integrated photon
flux from 2\,MeV to infinity of at least
10$^{10}$\,photons\,cm$^{-2}$\,s$^{-1}$
will be needed for at least 10$^9$ years in order to give a B/Be ratio lower
than 30.
Assuming the Galaxy to be a representative large spiral, we have found this
flux to be compatible with the observed $\gamma$-ray background
(cf. Prantzos \& Cass\'e and references therein), provided that the flux is
concentrated into objects with a total surface area on the order 1\,(ly)$^2$.
In order to convert enough carbon and oxygen into beryllium and boron it is
required that a fraction, $X$, of all matter in the halo gas is exposed to this
flux and that this material is efficiently mixed into the ISM prior to star
formation.
If the production of carbon and oxygen is not spatially strongly correlated
with
the $\gamma$-ray flux, $X$ is as great as 10$^{-4}$, which seems to lead to
very severe dynamical problems.
Even if the heavy target nuclei are produced by the $\gamma$ radiating objects
themselves, major dynamical problems would still remain.

We find these different alternatives to cosmic ray spallation improbable as
explanations of the Be and B abundances, while, on the other hand,
the observed B/Be ratio agrees very well with that
expected from cosmic ray spallation which supports this origin.
Additional support to the latter hypothesis, as compared with that of a
cosmological or other primordial but not that of a stellar origin, is given by
the linear relation between the B and O abundances found by
Gilmore et~al. (1992).
This linear relation is, however, not in itself trivial to understand in terms
of cosmic ray spallation (for different cosmic ray spallation scenarios, see
Duncan et~al. 1992, Pranzos et~al. 1993, Pranzos \& Cass\'e 1993, Pagel 1993,
Feltzing \& Gustafsson 1994).
We conclude, however, that cosmic ray
spallation is the probable origin of both Be and B in Population II stars.

Finally we note that the solar system meteoritic $N_{\rm B}/N_{\rm Be}$ ratio
$28\pm 4$ (Anders \& Grevesse 1989) is within our limits of uncertainty,
implying that the same process could in principle be responsible for the
production of B and Be throughout the history of the Galaxy.

Observations of Be and B abundances for stars with [Fe/H]$\;< -3.0$ and of
$^{11}$B/$^{10}$B ratios for Pop~II stars would give further important clues
for distinguishing between the different formation mechanisms, and also make it
possible to further constrain the cosmic-ray spallation hypothesis.
Current observational resources make both these goals marginally achievable.

\acknow{
It is a pleasure to thank Jeremy Walsh of the European Coordination
Facility (ECF) at Garching bei M\"unchen for his keen assistance with the data
reductions, generous sharing of his time and experience, and for valuable
comments to the manuscript.
We also thank Juliet Pickering, Imperial College, London, for providing
a figure of the line profile of the 2496.716\,\AA\ line.
BE and BG thank the Swedish National Space Board and the Swedish Natural
Science Research Council for support.
}

\begref{References}
\ref Boyd R.N., Fencl H.S. 1991, ApJ 373, 84
\ref Boyd R.N., Kajino T. 1989, ApJ 336, L55
\ref Duncan D.K. 1992, Goddard High Resolution Spectrograph Handbook
Version 3.0, Space Telescope Science Institute, Baltimore
\ref Duncan D.K., Lambert D.L., Lemke M. 1992, ApJ 401, 584
\ref Edvardsson B., Andersen J., Gustafsson B., Lambert D.L., Nissen P.E.,
Tomkin J. 1993, A\&A 275, 101
\ref Feltzing S., Gustafsson B. 1994, ApJ, in press
\ref Gilmore G., Edvardsson B., Nissen P.E. 1991, ApJ 378, 17
\ref Gilmore G., Gustafsson B., Edvardsson B., Nissen P.E. 1992, Nat 357, 379
\ref Jedamzik K., Fuller G.M., Mathews G.J., Kajino T. 1993, preprint (astro-ph
9312066)
\ref Kajino T., Boyd R.N. 1990, ApJ 359, 267
\ref Kawano L.H., Fowler W.A., Kavanagh R.W., Malaney R.A. 1991, ApJ 372, 1
\ref Kiselman D., 1994, A\&A, in press (astro-ph 9401018)
\ref Malaney R.A. 1992, ApJ 398, L45
\ref Malaney R.A., Fowler W.A. 1989, ApJ 345, L5
\ref Nissen P.E. 1993, Rev. Mex. Astron. Astrophys., in press
\ref Nissen P.E., Gustafsson B., Edvardsson B., Gilmore G. 1993, A\&A, in press
\ref Olive K.A., Prantzos N., Scully S., Vangioni-Flam E. 1993, preprint
(UMN-TH-1203/93)
\ref Pagel B.E.J. 1993, preprint (NORDITA 93/56A)
\ref Pickering J.C. 1993, Thesis, Blackett Laboratory, Imperial College, London
\ref Pickering J.C. 1994, private communication
\ref Prantzos N., Cass\'e M. 1993, preprint (Inst. d'Astrophysique de Paris
No. 429)
\ref Prantzos N., Cass\'e M., Vangioni-Flam E. 1993, ApJ 403, 630
\ref Ryan S.G., Norris J.E., Bessell M.S., Deliyannis C.P. 1992, ApJ 388, 184
\ref Steigman G., Walker T.P. 1992, ApJ 385, L13
\ref Terasawa N., Sato K. 1990, ApJ 362, L47
\ref Thomas D., Schramm D.N., Olive K.A., Mathews G.J., Meyer B.S, Fields B.D.
1993, ApJ, submitted (astro-ph 9308026)
\ref Vangioni-Flam E., Cass\'e M., Audouze J., Oberto Y. 1990, ApJ 364, 568
\ref Woosley S.E., Hartman D.H., Hoffman R.P., Haxton W.C. 1990, ApJ 356, 272
\endref
\end